\documentclass[leqno,a4paper,12pt]{article}
\usepackage[centertags]{amsmath}
\usepackage{amsfonts}
\usepackage{amssymb}
\usepackage{multirow}
\usepackage{amsthm}
\usepackage{pifont}
\usepackage[ansinew]{inputenc}
\usepackage{graphicx}

\theoremstyle{plain}
\newtheorem{thm}{Theorem}[]
\newtheorem{rem}{Remark}[]

\begin{document}

\title{Partial differential equation about the prevalence of a chronic disease in the
presence of duration dependency}
\author{Ralph Brinks}
\date{}
\maketitle

\begin{abstract}
The illness-death model of a chronic disease consists of the
states \emph{Normal, Disease,} and \emph{Death}. In general, the
transition rates between the states depend on three time scales:
calendar time, age and duration of the chronic disease. Previous
works have shown that the age-specific prevalence of the chronic
disease can be described by differential equations
\cite{Bri11,Bri12b} if the duration is negligible. This article
derives a partial differential equation (PDE) in the presence of
duration dependency. As an important application, the PDE allows
the calculation of the age-specific incidence from cross-sectional 
surveys.
\end{abstract}

\section{Introduction}
\begin{figure}[ht]
  \centering
  \includegraphics[keepaspectratio,width=0.85\textwidth]{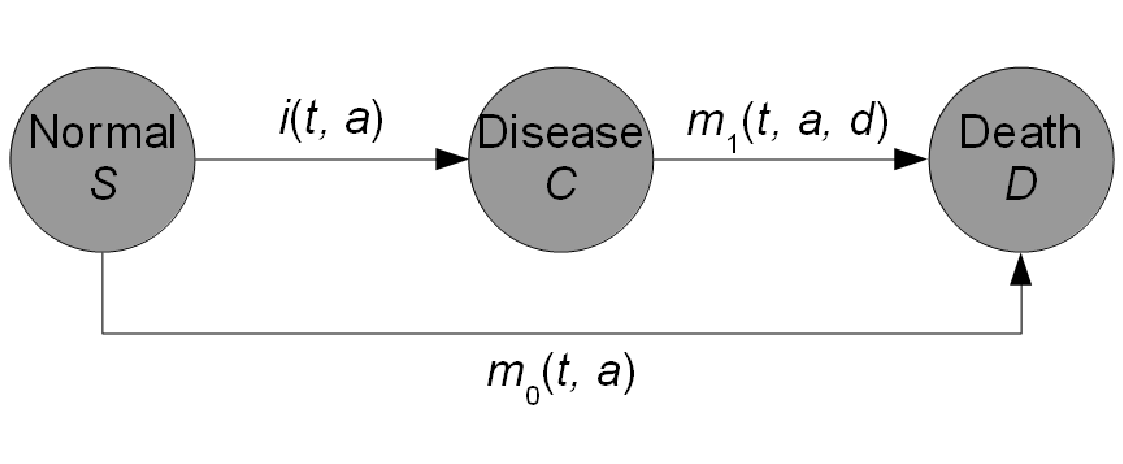}
\caption{Illness-death model with three states and the associated
transition rates. Persons in the state \emph{Normal} do not have
the chronic disease. In case of an onset, they change to the state
\emph{Disease}.} \label{fig:CompModel}
\end{figure}

The articles \cite{Bri11} and \cite{Bri12b} deal with the
illness-death model as shown in Figure \ref{fig:CompModel}. The
model describes a population consisting of healthy and ill persons
with respect to a chronic (i.e., irreversible) disease. Let the number
of persons in the states \emph{Normal} and \emph{Disease} be
denoted with $S$ and $C$. The transition rates between the states
are the incidence rate $i$ and the mortality rates $m_0$ and $m_1$
of the healthy and the diseased persons, respectively. In general,
these rates depend on the calendar time $t$, on the age $a$ and in
case of $m_1$ also on the duration $d$.

The articles \cite{Bri11} and \cite{Bri12b} derived differential
equations (DEs) about the age-specific prevalence $p =
\tfrac{C}{C+S}$ and the rates $i$, $m_0$ and $m_1:$

\begin{equation}\label{e:structure}
\partial p = (1-p) \, \left ( i - p \left ( m_1 - m_0 \right ) \right
).
\end{equation}
The symbol $\partial$ means a differential operator, either 
$\partial = \tfrac{\mathrm{d}}{\mathrm{d}a}$ or $\partial = 
\tfrac{\partial}{\partial t} + \tfrac{\partial}{\partial a},$ 
depending on the setting.

\bigskip

The advantages of an equation of the form \eqref{e:structure} is
the relatively simple relation between the rates $i, m_0, m_1$ 
and the
prevalence $p$. If the rates on the right hand side of the
equation are known, the DE can be solved for $p$. This is called
the \emph{direct problem} \cite{Bri12a}. Furthermore, if $p, m_0$ and
$m_1$ are known, Equation
\eqref{e:structure} can be solved for the incidence $i$. This
\emph{inverse problem} allows the calculation of the incidence of
a chronic disease from cross-sectional data. Usually in epidemiology, 
incidence
rates are surveyed by lengthy (and thus expensive) follow-up
studies. The possibility to use less complex cross-sectional data
instead of follow-up studies is an enormous benefit and has been requested 
for a long time (see \cite{Hen10} for an overview).

\bigskip

The DEs in \cite{Bri11,Bri12b} were restricted to the case of
$m_1$ being independent from the disease duration $d$. For some
chronic diseases, this is not the case. An example is type 2 diabetes: 
after physiological onset, most patients do not have any symptoms and the
disease remains undetected. After onset of clinical symptoms, the
mortality already $m_1$ is higher compared to the mortality $m_0$ of a
healthy person with the same age (and sex). Then, the therapy starts. 
During therapy, the mortality decreases and after some time with the disease the
mortality increases again due to late sequelae. In the Danish
diabetes register the J-shaped mortality ratio (i.e.,
$\tfrac{m_1}{m_0}$) is empirically evident \cite{Car08}. There are
other chronic diseases with duration being an important factor for
mortality, for example, dementia \cite{Rai10} and systemic lupus
erythematosus \cite{Ber06}. It is likely that duration is an
important covariate for many of chronic diseases.

The aim of this article is the derivation of a partial
differential equation (PDE) similar to Equation
\eqref{e:structure} in case the mortality rate $m_1$ depends on
the duration $d$. Thus, this work is a generalization of the the
previous articles \cite{Bri11,Bri12b}.


\section{Basic equations for the illness-death model}
As in \cite{Bri12b} let $S(t, a)$ and $C(t, a)$ denote the numbers
of healthy and diseased persons aged $a$ at time $t$. Based on the
possible transitions from the \emph{Normal} state in Figure
\ref{fig:CompModel} we get the following Cauchy problem for $S(t,
a)$:
\begin{eqnarray}
(\partial_t + \partial_a) \, S(t, a) &=& - \left ( m_0(t, a) + i(t, a) \right )
\, S(t, a) \label{e:PDE_S_ta} \\
S(t - a, 0) &=& S_0(t - a). \nonumber
\end{eqnarray}

Here $S_0(t - a) = S(t-a, 0)$ is the number of healthy newborns at
calendar time $t-a$. Henceforth, the notation $\partial_x$ means
the partial derivative for the variable $x, ~x \in \{t, a, d\}.$
The rates $i, m_0$ and $m_1$ in Figure \ref{fig:CompModel} are
assumed to be sufficiently smooth, such that the common uniqueness
and existence theorems hold \cite{Pol00}.

The unique solution $S(t, a)$ is
\begin{equation}
S(t, a) = S_0(t - a) \, \exp \left ( - \int_0^a m_0(t-a+\tau, \tau) +
i(t-a+\tau, \tau) \, \mathrm{d}\tau \right ) \label{e:S}.
\end{equation}

\bigskip

In case of the diseased persons, we have to distinguish between
different disease durations. Let $C^\star(t, a, d)$ be the number
of diseased persons aged $a$ at time $t$ having the disease for
the exact duration $d, ~d < a.$

For $C^\star(t, a, d)$ we also have a PDE:
\begin{equation}
  (\partial_t + \partial_a + \partial_d) \, C^\star(t,a,d) =
   - C^\star(t, a, d) \, m_1(t, a, d). \label{e:PDE_tad}
\end{equation}

The initial condition for making Equation \eqref{e:PDE_tad} a Cauchy problem 
stems from
the assumption that newly diseased persons may only enter from the
\emph{Normal} state. This means: $C^\star(t, a, 0) = i(t, a) \, S(t, a)$ for
all $t, a.$

The unique solution of the Cauchy problem for $C^\star(t, a, d)$
is:
\begin{eqnarray*}
   C^\star(t, a, d) &=& C^\star(t-d, a-d, 0) \, \exp
                        \left ( - \int_0^d m_1(t-d+\tau, a-d+\tau, \tau) \, \mathrm{d}\tau
                        \right )\\
                    &=& i(t-d, a-d) \,  S(t-d, a-d) \,
                        e^{- \int\limits_{0}^{d} m_1(t-d+\tau, a-d+\tau, \tau) \,
                        \mathrm{d}\tau}.
\end{eqnarray*}

The total number $C(t, a)$ of diseased persons can be obtained by
integration:
\begin{eqnarray}
C(t, a) &=& \int_0^a C^\star(t, a, \delta) \, \mathrm{d}\delta \nonumber \\
        &=& \int_0^a i(t-\delta, a-\delta) \,  S(t-\delta, a-\delta) \,
            e^{- \int\limits_{0}^{\delta } m_1(t-\delta+\tau, a-\delta+\tau, \tau)
            \mathrm{d}\tau} \mathrm{d}\delta \label{e:C}
\end{eqnarray}

\bigskip

Summing up the previous equations, we get the following result for
the age-specific prevalence. The result was given without
derivation in \cite{Kei91}.
\begin{thm}
For the age-specific prevalence $$p(t, a) = \tfrac{C(t, a)}{S(t,
a) + C(t, a)}$$ it holds
\begin{equation}
p(t, a) = \frac{\int\limits_0^a i(t-\delta, a-\delta) \, \mathcal{M}_{t,a}(a-\delta) \,
          e^{- M_1(t, a, \delta)} \, \mathrm{d}\delta}
          {\mathcal{M}_{t,a}(a) + \int\limits_0^a i(t-\delta, a-\delta) \, \mathcal{M}_{t,a}(a-\delta) \,
          e^{- M_1(t, a, \delta)} \, \mathrm{d}\delta}, \label{e:p}
\end{equation}
where
\begin{equation*}
\mathcal{M}_{t, a}(y) := \exp \left ( -\int_0^y m_0(t-a+\tau,
\tau) + i(t-a+\tau, \tau) \mathrm{d}\tau \right )
\end{equation*}
and
$$M_1(t, a, d) := \int_{0}^{d} m_1(t - d + \tau, \,a - d + \tau , \,\tau ) \,
\mathrm{d}\tau.$$
\end{thm}

\bigskip

As mentioned above, Equation \eqref{e:p} allows the calculation of
the age-specific prevalence if the rates $i(t, a), m_0(t, a)$ and
$m_1(t, a, d)$ are known.

\bigskip

As an observation we get:
\begin{rem}
$p$ does not depend on the number of newborns $S_0$.
\end{rem}

To conclude this section, we list the assumptions that have
been used to deduce Equation \eqref{e:p}:
\begin{enumerate}
    \item We consider a chronic (i.e., irreversible) disease. This means,
    there is no transition from the \emph{Disease} to the \emph{Normal} state.
    \item The transitions are described by the rates\footnote{In stochastic
    contexts, what here is called \emph{rate} is synonymously denoted as \emph{density}.}
    $i, m_0$ and $m_1$. The rates are sufficiently smooth.
    \item Newborns are disease-free.
    \item Except for disease-free newborns, no one enters the
    population.
    \item The only way out of the population is the \emph{Death}
    state.
\end{enumerate}

\section{Differential equations for the age-specific prevalence}
An enormous drawback of Equation \eqref{e:p} lies in the fact that
it cannot be solved for the incidence $i$ even if the mortality
rates $m_0$ and $m_1$ are known. If the duration $d$ does not play
a role, it can be shown that $p$ fulfills a simple differential
equation of the form \eqref{e:structure} \cite{Bri11,Bri12b}. In
case of independence from $d$, we can also release some of the
assumptions given at the end of the previous section (see
\cite{Bri12b} for details).

In this section, it is shown that with the assumptions recapped 
above, the age-specific prevalence $p$
fulfills a PDE similar to Equation \eqref{e:structure}. For this,
define the mortality rate $m_1^\star$ by:
\begin{equation}\label{e:mstar}
 m_1^\star(t, a) := \frac{\int\limits_0^a m_1(t, a, \delta) \, C^\star(t, a, \delta) \mathrm{d}\delta}
                       {\int\limits_0^a C^\star(t, a, \delta)
                       \mathrm{d}\delta}.
\end{equation}
Obviously, this definition needs the assumption that 
$\int_0^a C^\star(t, a, \delta) \mathrm{d}\delta (= C(t, a)) \neq 0$ 
for $(t, a).$ At points $(t, a)$ such that $C(t, a) = 0$ 
define $m_1^\star(t, a) := 0.$

Since the numerator in Equation \eqref{e:mstar} is the total
number of ``incident'' death cases among all diseased persons aged
$a$ at $t$, the rate $m_1^\star(t, a)$ is the \emph{overall
mortality rate} of the diseased persons. 

\bigskip

Assumed that $C$ fulfilled the PDE
\begin{equation}\label{e:Cmstar}
    (\partial_t + \partial_a) \, C(t, a) = - m_1^\star(t, a) \, C(t, a) + i(t, a) \, S(t,a),
\end{equation}
then for $p = \tfrac{C}{S+C}$ it would hold
\begin{equation}\label{e:pmstar}
(\partial_t + \partial_a) \, p = (1-p) \, \bigl ( i - p \left (
m_1^\star - m_0 \right ) \bigr ).
\end{equation}

\begin{proof}
Let $\partial := \partial_t + \partial_a.$ Application of the
quotient rule to $p = \tfrac{C}{S+C}$ yields $\partial p =
\frac{(1-p) \, \partial C - p \, \partial S}{S+C}.$ Inserting
Equations \eqref{e:PDE_S_ta} and \eqref{e:Cmstar} into the
nominator of this expression yields Equation \eqref{e:pmstar}.
\end{proof}

It remains open to prove Equation \eqref{e:Cmstar}. With $\partial
= \partial_t + \partial_a$ it holds:
\begin{eqnarray*}
\partial C(t, a)
    &=& \partial \int\limits_0^a C^\star(t, a, \delta) \, \mathrm{d}\delta  \\
    &=&  \int\limits_0^a \partial C^\star(t, a, \delta) \, \mathrm{d}\delta + C^\star(t, a, a)\\
    &=&  \int\limits_0^a (\partial_t + \partial_a + \partial_d) C^\star(t, a, \delta) \, \mathrm{d}\delta - \int\limits_0^a \partial_d C^\star(t, a, \delta) \, \mathrm{d}\delta + C^\star(t, a, a)\\
    &=& - \int\limits_0^a m_1(t, a, \delta) \, C^\star(t, a, \delta) \, \mathrm{d}\delta - \int\limits_0^a \partial_d C^\star(t, a, \delta) \, \mathrm{d}\delta + C^\star(t, a, a)\\
    &=& - m_1^\star(t, a) \, C(t, a) - \bigl ( C^\star(t, a, a) - C^\star(t, a, 0) \bigr ) + C^\star(t, a, a) \\
    &=& - m_1^\star(t, a) \, C(t, a) + i(t, a) \, S(t, a).
\end{eqnarray*}
For the second equality, Leibniz's integral rule (also known as
\emph{differentiation under the integral sign}) has been used.

\bigskip

In summary, we get:
$$(\partial_t + \partial_a) \, p = (1-p) \, \bigl ( i - p\, (m_1^\star - m_0) \bigr).$$

\begin{rem}\label{r:m}
If $m_1$ does not depend on the duration $d$, it holds
$m_1^\star(t, a, \delta) = m_1^\star(t, a) = m_1(t, a).$ In this case,
Equation \eqref{e:pmstar} is equal to the PDE $(\partial_t +
\partial_a) \, p = (1-p) \, ( i - p\, (m_1 - m_0))$ as in
\cite{Bri12b}.
\end{rem}

\section{About the overall mortality $m_1^\star$ of the diseased}
In this section, the overall mortality $m_1^\star$ of the diseased
is examined further and illustrated by a numerical example.

For $C(t, a) > 0$ it holds 
$$m^\star_1(t, a) = \int\limits_0^a m_1(t, a, \delta) \, \frac{C^\star(t, a, \delta)}{C(t, a)} \mathrm{d}\delta.$$ 
The fraction $\tfrac{C^\star(t, a, \delta)}{C(t, a)}$ is the duration distribution of the diseased persons at 
$(t, a)$. Thus, $m^\star_1(t, a)$ depends on the duration distribution of the disease. 

\bigskip

In addition, by inserting the expressions for $C^\star(t,a,d)$ into Equation \eqref{e:mstar} and
cancelling out $S_0(t-a)$ one obtains
\begin{equation}\label{e:mstar2}
 m_1^\star(t, a) = \frac{\int\limits_0^a m_1(t, a, \delta) \, i(t-\delta, a-\delta) \, \mathcal{M}_{t,a}(a-\delta) \,
          e^{- M_1(t, a, \delta)} \, \mathrm{d}\delta}
                       {\int\limits_0^a i(t-\delta, a-\delta) \, \mathcal{M}_{t,a}(a-\delta) \,
          e^{- M_1(t, a, \delta)} \, \mathrm{d}\delta}.
\end{equation}

Hence, the overall mortality rate of the diseased depends on the age-specific incidence rate $i.$ 
To be more specific, Equation \eqref{e:mstar2} shows that at the point $(t,a)$, the incidence $i$ 
at all the previous points in time $(t-\delta, a-\delta), ~\delta \in [0, a],$ contributes to 
$m^\star_1(t, a).$ The dependence of $m^\star_1(t, a)$ on the past incidence 
$i(t-\delta, a-\delta), ~\delta \in [0, a],$ is not surprising. 
Consider two chronic diseases with onset at ages $a_\ell, ~\ell = 1, 2,$ i.e., $i_\ell(t, a) = 0$ 
for all $(t, a)$ where $a \le a_\ell.$ If, say, disease 1 has a later onset than disease 2, $a_1 > a_2,$ 
then by definition it holds $m^\star_{1, i_1}(t, a') = 0$ for all $a' \in [a_2, a_1],$
 whereas $m^\star_{1, i_2}(t, a')$ may be greater than 0.

\bigskip

To illustrate the dependency of $m^\star_1$ on the incidence we set up an example. We assume two 
hypothetical chronic diseases in a fictional population.
Henceforth, we assume that $t, a$ and $d$ are measured in years $t, a, d \ge 0.$

\smallskip

For the mortality rate $m_0$ of the normal population we assume the 
Strehler-Mildvan form $m_0(t, a) = \exp(-11-0.04 \, t+0.1 \, a).$ This is an 
approximation for the general mortality in the male German population as used by the official 
population projection of the Federal Statistical Office \cite{Fed09}.
The calendar time $t$ is measured in years since 2010.

\smallskip

For the incidence of the first disease, we assume $i_1(t, a) = i_1(a) = \tfrac{(a – 30)_+}{3000},$ 
where $x_+$ means $x_+ = \max(0, x).$ The incidence of the second disease is just
$i_2 = 0.1 \, i_1.$ 

In addition, let the mortality $m_1$ be given by 
$m_1(t, a, d) =  R(d) \, m_0(t, a),$ where $R(d) = (0.2 \, d - 1)^2 + 1.$

\bigskip

\begin{figure}[ht]
  \centering
  \includegraphics[keepaspectratio,width=0.85\textwidth]{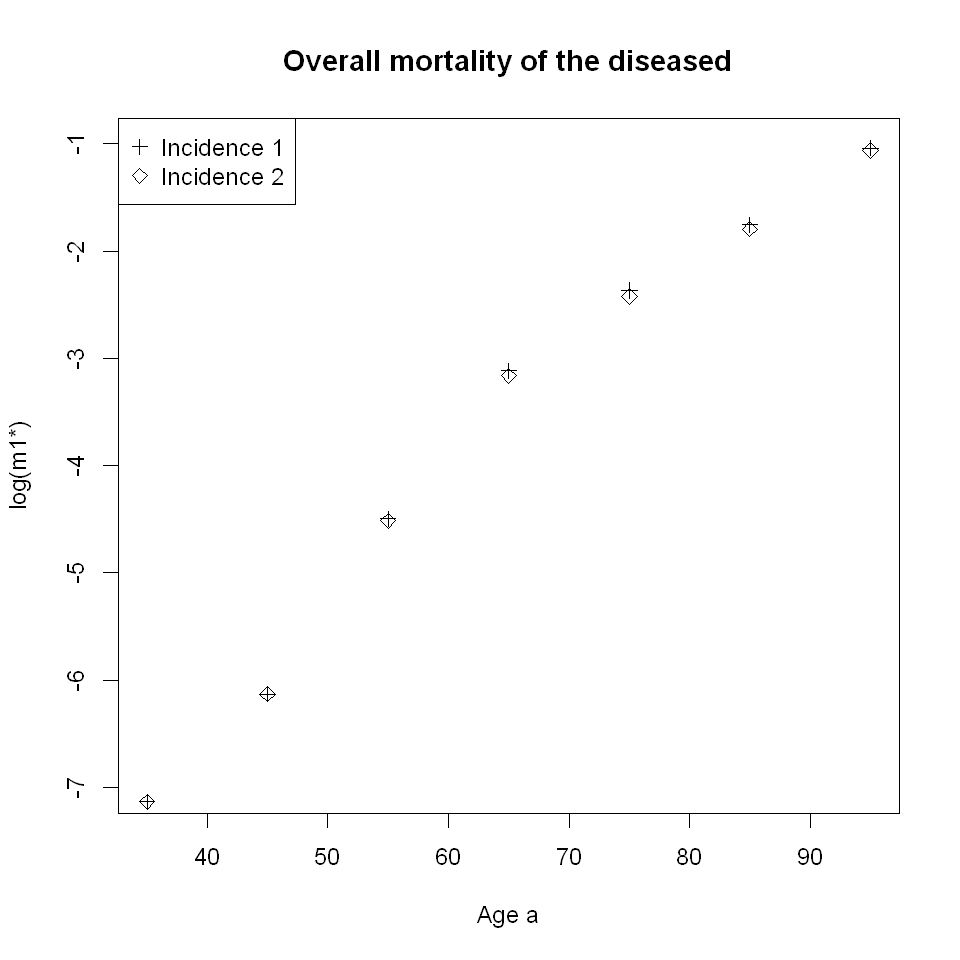}
\caption{Logarithm of the overall mortality of the diseased for two different incidence rates.}
\label{fig:Results}
\end{figure}

Figure \ref{fig:Results} shows the overall mortality $m^\star_{1, i_\ell}(t, a)$ in year $t=1$ 
for the different incidences $i_\ell, \ell = 1, 2,$ over the age $a$. The values have been 
calculated by Equation \eqref{e:mstar2} using Romberg's rule for integration. For ages below 50 
the values $m^\star_{1, i_\ell}(t, a)$ rather agree, but diverge for ages between 50 and 80. For higher 
ages, the difference between $m^\star_{1, i_1}$ and $m^\star_{1, i_2}$ decreases again.

\begin{figure}[ht]
  \centering
  \includegraphics[keepaspectratio,width=\textwidth]{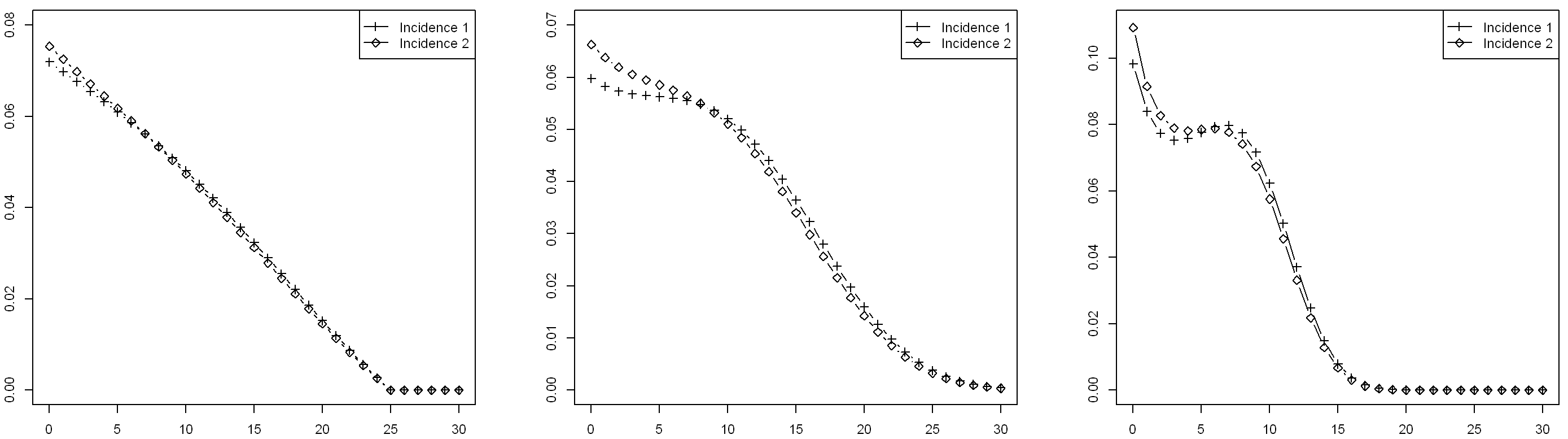}
\caption{Duration distribution of the diseased persons for the different incidence rates 
$i_\ell, ~\ell = 1$ (crosses)  and $\ell = 2$ (diamonds), at different ages: $a=55$ (left), $a=75$ (middle), $a=95$ (right).}
\label{fig:DurDist}
\end{figure}

The reason for the differences can be seen in Figure \ref{fig:DurDist}, which depicts the duration 
distribution of the diseased persons $C(t=1, a)$ for ages $a = 55, 75, 95.$

From Figures \ref{fig:Results} and \ref{fig:DurDist} it becomes apparent that, although 
$m_1(t, a, d)$ are the same for both diseases, the overall mortality $m_1^\star$ depends on the 
duration distribution of the diseased persons, which in turn depends on the incidence.

\section{Discussion}
This work generalizes the results of the previous articles \cite{Bri11} and \cite{Bri12b}
for the illness-death model as shown in Figure \ref{fig:CompModel}. In contrast to the previous
articles, we do not need the assumption that the mortality of the diseased persons is 
independent from the duration of the disease. The modification that arises from possible duration dependence 
is the introduction of a measure for the mortality of the diseased persons, the overall mortality 
$m^\star_1.$ We have shown that $m^\star_1$ depends on the duration distribution of the diseased 
persons, which in turn is a consequence of the incidence rate. With respect to the aim of deriving 
the age-specific incidence from prevalence data, this seems to be a drawback. However, if the 
duration distribution of the diseased persons is known, then the calculation of $m^\star_1$ does 
not impose a problem. The request for having knowledge about the duration distribution is not unusual. 
It arises from the fundamental demand of epidemiology that the sample population (in a survey) should
be a representative subset of the target population. Given that the duration distribution is known, 
calculation of $m^\star_1$ from the duration dependent mortality $m_1$ is possible. Another way of 
obtaining $m_1^\star$ is surveying it directly in an epidemiological study. If the sample population
is representative for the target population -- this means also representative with respect to the duration distribution -- $m_1^\star$ may be obtained from standard survival analysis if the duration dependency is ignored. In this way, the overall mortality $m^\star_1$ is an easily accessible epidemiological measure.
 
If, furthermore, the mortality $m_0$ of the healthy, or at least the general mortality 
$m = p \, m^\star_1 + (1-p) \, m_0$ is known, then Equation \eqref{e:pmstar} can be used to obtain 
the incidence rate $i$.


{}

\end{document}